\documentclass[showpacs,preprintnumbers,amsmath,amssymb,aps,prb,floatfix,reprint,superscriptaddress]{revtex4-1}

\usepackage{graphicx}
\usepackage[colorlinks=true,citecolor=blue,linkcolor=blue,urlcolor=blue]{hyperref}
\usepackage{xcolor,soul}
\DeclareMathOperator{\atantwo}{atan2}

\begin{document}

\title{Tunable Snell's law for spin waves in heterochiral magnetic films}
\author{Jeroen Mulkers}
\email[Email: ]{jeroen.mulkers@uantwerpen.be}
\affiliation{Departement Fysica, Universiteit Antwerpen, Groenenborgerlaan 171, B-2020 Antwerpen, Belgium}
\affiliation{DyNaMat Lab, Department of Solid State Sciences, Ghent University, Ghent, Belgium}
\author{Bartel Van Waeyenberge}
\affiliation{DyNaMat Lab, Department of Solid State Sciences, Ghent University, Ghent, Belgium}
\author{Milorad V. Milo\v{s}evi\'c}
\email[Email: ]{milorad.milosevic@uantwerpen.be}
\affiliation{Departement Fysica, Universiteit Antwerpen, Groenenborgerlaan 171, B-2020 Antwerpen, Belgium}
\date{\today}

\begin{abstract}
Thin ferromagnetic films with an interfacially-induced Dzyaloshinskii-Moriya interaction (DMI) exhibit non-trivial asymmetric dispersion relations that lead to unique and useful magnonic properties. Here we derive an analytical expression for the magnon propagation angle within the micromagnetic framework and show how the dispersion relation can be approximated with a comprehensible geometrical interpretation in the $k$-space of the propagation of spin waves. We further explore the refraction of spin waves at DMI interfaces in heterochiral magnetic films, after deriving a generalized Snell's law tunable by an in-plane magnetic field, that yields analytical expressions for critical incident angles. The found asymmetric Brewster angles at interfaces of regions with different DMI strengths, adjustable by magnetic field, support the conclusion that heterochiral ferromagnetic structures are an ideal platform for versatile spin-wave guides.
\end{abstract}

\pacs{75.30.Ds,75.70.Ak,75.70.Cn,75.78.Cd}
\keywords{Dzyaloshinskii-Moriya interaction, micromagnetism, perpendicular magnetic anisotropy, magnons, spin waves, refraction}

\maketitle

\section{Introduction}

Spin waves (and their quasiparticle counterpart, magnons), the collective excitations in magnetic spin systems coupled by exchange interactions, present a wide variety of unique properties and prospective applications that continuously inspire fundamental research. Just like any wave, spin waves experience dispersion caused either by geometric boundary or by interaction with the transmitting medium. The Dzyaloshinskii-Moriya interaction (DMI), present in magnetic materials with broken inversion symmetry \cite{Dzyaloshinsky1958,Moriya1960,Dzyaloshinsky1964,Crepieux1998}, has a chiral character and introduces an asymmetry in the spin-wave dispersion relation \cite{Udvardi2009,Costa2010,Cortes-Ortuno2013,Moon2013}. This leads to a plethora of remarkable phenomena such as the asymmetric frequency shift measured in spin-polarized electron-energy-loss and Brillouin light-scattering experiments \cite{Zakeri2010,Di2015,Nembach2015,Cho2015,Belmeguenai2015,Belmeguenai2016,Stashkevich2015,Lee2016}, the magnon Hall effect \cite{Onose2010,Murakami2017}, a non-trivial spin-wave power flow and unidirectional caustic beams \cite{Kim2016a}, unidirectional spin-wave emitters \cite{Bracher2017}, nonreciprocal spin-wave channeling along spin textures \cite{Garcia-Sanchez2014,Garcia-Sanchez2015}, and a non-trivial refraction of spin waves at domain walls \cite{Yu2016}, to name a few. Yet, chiral magnonics is still believed to be at the doorstep of its full potential.

Over the last years, immense experimental progress was made with layered heterostructures, where DMI is interfacially induced \cite{Woo2016,Moreau-Luchaire2016}. It motivated the exploration of heterochiral structures - films in which DMI can be spatially varied via engineering of the substrate and/or the capping layer \cite{Balk2017,Wells2017}. Such structures have been already predicted to strongly confine magnetic skyrmions \cite{Mulkers2017} and increase their lifetime \cite{Stosic2017}, both essential for skyrmionic devices. In this paper we take the next step and examine the propagation of spin waves in heterochiral films with a spatially-engineered DMI and perpendicular magnetic anisotropy. Starting from the dispersion relation, we derive an analytical expression for the magnon propagation angle in monochiral films. Next, we show how the non-trivial dispersion relation can be approximated by circular isofrequencies to provide a comprehensible geometrical interpretation in $k$-space. This can then be conveniently used to understand the refraction at interfaces where micromagnetic parameters change, such as in a heterochiral magnetic film. We go on to derive the generalized Snell's law for spin waves at interfaces where DMI changes, broadly tunable by in-plane magnetic field. Although our derived relation is unique to chiral magnetic interfaces, it has similar consequences as found in metamaterials for photonics and phononics \cite{Yu2011,Sun2012,Sun2012a,Yu2014,Chen2012,Li2014,Zhu2015}, and thus bears general relevance to wave propagation in (hetero)chiral media.

The paper is organized as follows. In Sec. II, we outline the theoretical framework of our calculations. Sec. III is devoted to description of propagation of spin-waves in magnetic films with homogeneous DMI. The prime topic of the paper, the spin-wave refraction at interfaces where DMI changes, is addressed in Sec. IV. Our results are summarized in Sec. V.

\section{Micromagnetic framework}

We describe the magnetization of a ferromagnetic film by a 2D continuous field~$\vec{M}(x,y) = M_{\text{s}} \vec{m}(x,y)$ with a constant magnetization modulus~$|\vec{M}|=M_{\text{s}}$ and magnetization direction~$\vec{m}(x,y)$. The dynamics of the magnetization are governed by the Landau-Lifshitz-Gilbert~(LLG) equation
\begin{equation}
    \vec{m}_t = \frac{-\gamma}{1+\alpha^2} \left( \vec{m} \times \vec{H}_{\text{eff}} + \alpha \left[ \vec{m} \times( \vec{m} \times \vec{H}_{\text{eff}}) \right] \right),\label{eq:llg}
\end{equation}
with gyromagnetic ratio~$\gamma$ and damping factor~$\alpha$. At each point in the film, the magnetization precesses around the effective magnetic field, which is the functional derivative of the magnetic free energy~$E=\int\varepsilon \text{d} V$ with respect to the magnetization: $\vec{H}_{\text{eff}} = -\delta E/\delta\vec{M}$.

The local energy density~$\varepsilon(x,y)$ of a given magnetization~$\vec{M}$ has multiple sources, and we consider the following: exchange, perpendicular anisotropy, Zeeman interaction due to an in-plane applied field, DMI, and demagnetization. We focus on the propagation of spin waves in ultrathin films ($<$1.5~nm), for which the occurrence of unidirectional caustic beams becomes negligible \cite{Kim2016a} and for which we can approximate the demagnetization field via an effective anisotropy~$K_{\text{eff}} =K-1/2\mu_0 M^2_{\text{s}}$.
The expressions for the remaining energy-density terms are, respectively,
\begin{align}
    \varepsilon_{\text{ex}}  &=  A [ \left(\partial_x \vec{m}\right)^2 + \left(\partial_y\vec{m}\right)^2], \label{eq:exchange} \\
    \varepsilon_{\text{anis}}&= -K_{\text{eff}}m_z^2,\\
    \varepsilon_{\text{ext}} &= -\vec{B} \cdot \vec{m} M_{\text{s}}, \\
    \varepsilon_{\text{dmi}} &= D
    \begin{aligned}[t]
        &  [ m_x\partial_x m_z - m_z\partial_x m_x \\
        & \quad + m_y \partial_y m_z-m_z\partial_y m_y ], \label{eq:dmi}\\
    \end{aligned}
\end{align}
with exchange stiffness $A$, DMI strength $D$, effective anisotropy constant $K_{\text{eff}}$, and a bias field $\vec{B}\perp \hat{e}_z$.
To simplify the notation, we introduce the exchange length~$\xi=\sqrt{A/K_{\text{eff}}}$ and the critical DMI strength~$D_{\text{c}} =4\sqrt{AK_{\text{eff}}}/\pi$ \cite{Rohart2013}. In this paper we only consider first-order deviations from a uniformly magnetized film, which is the ground state for DMI strengths below $D_{\text{c}}$.

Due to the perpendicular anisotropy, the magnetic moments are parallel to the normal of the film ($z$-axis). However, applying an in-plane magnetic field~$\vec{B}=B(\cos\beta,\sin\beta,0)$ will tilt the magnetic moments in the direction of the applied field. This tilting of the magnetic moments is necessary to observe first-order effects of DMI on spin waves. The relaxed uniform magnetization is given by $\vec{m}_0 = (\cos\beta\sin\theta,\sin\beta\sin\theta,\cos\theta)$, with tilting angle
\begin{equation}
    \theta = \begin{cases}
        \arcsin \frac{M_{\text{s}} B}{2 K_{\text{eff}}} & \text{if } B\le B_{\text{c}},\\
        \pi/2 & \text{if } B\ge B_{\text{c}},
    \end{cases}\label{eq:theta}
\end{equation}
which is derived by minimizing the free energy, assuming a uniform magnetization. The magnetic moments are fully aligned with the in-plane magnetic field if its magnitude exceeds the critical value~$B_{\text{c}}=2K_{\text{eff}}/M_{\text{s}}$.

\section{Spin-wave propagation in monochiral films}

\subsection{Dispersion relation}

In order to derive the spin-wave dispersion relation, we study the time evolution [Eq.~\eqref{eq:llg}] of the first-order deviations from the equilibrium configuration~$\vec{m}_0$, omitting the damping term ($\alpha=0$), similarly to Refs.~\cite{Cortes-Ortuno2013,Moon2013}. With details of the derivation given in the supplemental material, the obtained spin-wave dispersion relation reads
\begin{equation}
    \frac{\omega}{\omega_{\perp}}
    = \sqrt{\left(\xi^2k^2 + \mathcal{B} - \sin^2\theta\right)\left(\xi^2k^2+\mathcal{B}\right)} -2\xi^2\vec{k}\cdot \vec{k}_0, \label{eq:dispersion}
\end{equation}
with $\omega_{\perp}= 2 \gamma K_{\text{eff}}/M_{\text{s}}$, $\mathcal{B}=\max(1,B/B_{\text{c}})$, and
\begin{equation}
    \vec{k}_0 = \frac{2 \sin\theta}{\pi\xi}\frac{D}{D_{\text{c}}} (\hat{e}_B\times\hat{e}_z).
\end{equation}
$\omega_{\perp}$ is the frequency of the precession of the magnetic moments around the anisotropy axis, in absence of other magnetic interactions. $\mathcal{B}$ is introduced to combine the two cases $B<B_{\text{c}}$ and $B>B_{\text{c}}$ in a concise mathematical expression.

The DMI, in combination with an applied in-plane field, causes a term linear in $\vec{k}$. This asymmetry introduces non trivial spin-wave phenomena. For example, as already known, it explains the frequency shift $\Delta \omega = |\omega(\vec{k})-\omega(-\vec{k})|$ measured in Brillouin light scattering measurements. This linear term also has an important influence on the propagation direction of spin-wave packets, and the refraction of spin waves at DMI interfaces, which is the main topic of this paper.

\subsection{Geometric interpretation}

The influence of the dispersion relation in Eq.~\eqref{eq:dispersion} on the propagation of spin waves is not easy to grasp intuitively. It is therefore useful to approximate the dispersion relation with circular isofrequencies in $k$-space, which can be done if $x=\sin^2\theta/(\xi^2k^2+\mathcal{B})$ is small. Note that the condition $0<x<1$ is always met and the Maclaurin series of functions of $x$ will yield good approximations in case of weak applied fields, strong applied fields, or small wavelengths.
The dispersion relation approximated with isofrequencies reads
\begin{equation}
    \frac{\omega}{\omega_{\perp}} \approx \frac{\omega_0}{\omega_{\perp}} + \xi^2 (\vec{k}-\vec{k}_0)^2, \label{eq:approx}
\end{equation}
with the minimal frequency
\begin{equation}
    \frac{\omega_0}{\omega_{\perp}} = \mathcal{B} - \frac{\sin^2\theta}{2} - \xi^2 \vec{k}_0^2,
\end{equation}
obtained when $\vec{k}=\vec{k}_0$. The vector pointing to the center of the circular isofrequencies $\vec{k}_0$ is independent of the frequency~$\omega$. It is also perpendicular to the magnetic field~$\vec{B}$, and proportional to DMI strength $D$ and the magnetic field, more precisely $\sin\theta$. The radius of the circular isofrequency $k_{\text{g}}$ depends on the frequency as
\begin{equation}
    \xi k_{\text{g}} = \sqrt{\frac{\omega-\omega_0}{\omega_{\perp}}}.
\end{equation}
Using this approximation, it becomes very easy to study the propagation of spin waves geometrically, as well as to examine the refraction of spin waves at interfaces where DMI changes.

\subsection{Magnon propagation angle}

The group velocity can be calculated exactly for the dispersion relation given in Eq.~\eqref{eq:dispersion}:
\begin{equation}
    \vec{v}_{\text{g}} = \nabla_k \omega = 2\omega_{\perp}\xi^2\left(\iota\vec{k}-\vec{k}_0\right)
    \quad \text{with} \quad
    \iota = \frac{1-\frac{1}{2}x}{\sqrt{1-x}}. 
    \label{eq:velocity}
\end{equation}
The propagation direction is always perpendicular to the isofrequencies in $k$-space. For the approximated dispersion relation, this means that the propagation direction is parallel to $\vec{k}-\vec{k}_0$, which corresponds to the exact solution for $\iota\approx 1$.

\begin{figure}[t]
    \centering
    \includegraphics{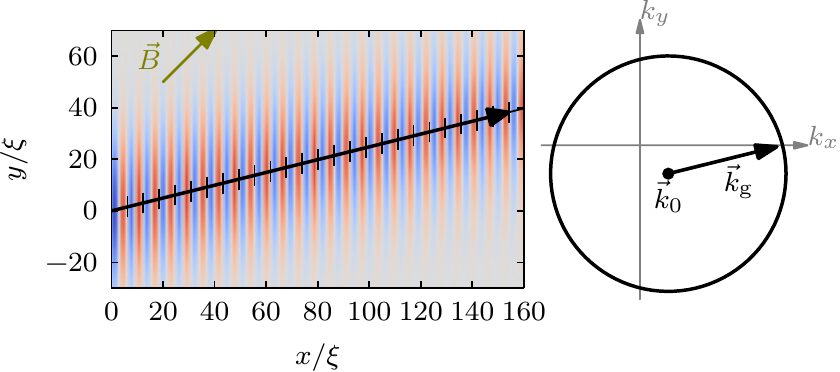} 
    \caption{Demonstration of the misalignment between the propagation direction of a wave packet and its wave vector for an in-plane field $B=0.5K_{\text{eff}}/M$ and $\beta=\pi/2$, DMI strength $D=0.9D_{\text{c}}$ and frequency $\omega=1.5\omega_{\perp}$. A Gaussian spin wave packet is artificially created at the left boundary and propagates to the right. The colors show the deviation from the equilibrium magnetization calculated by a full micromagnetic simulation. The black lines are the analytical predictions of the propagation direction and the wavefronts.}
    \label{fig:hall}
\end{figure}

In general, the propagation direction~$\iota \vec{k}-\vec{k}_0$ is not parallel to the wave vector~$\vec{k}$. It is trivial to prove that the angle between the wave vector~$\vec{k}=k(\cos\phi_k,\sin\phi_k)$ and the propagation direction is given by:
\begin{equation}
    \phi_{\text{prop}} = \atantwo\left(k_0\cos(\beta-\phi_k),\iota k-k_0\sin(\beta-\phi_k)\right).
\end{equation}
If $\iota k>k_0\sin(\beta-\phi_k)$, then the propagation direction has a component in the opposite direction of the wave vector $k$, hence the use of the $\atantwo$ function. This expression is useful when positioning an antenna to create spin waves with a desired propagation direction. The propagation angle (for a given direction of the wave vector $\vec{k}$) depends on the magnitude of the wave vector $|\vec{k}|$. This means that spin waves with the same wave vector direction, but different frequencies, propagate in different directions. Fig.~\ref{fig:hall} shows the results of a micromagnetic simulation using MuMax$^3$ \cite{Vansteenkiste2014} of the propagation of a Gaussian spin-wave packet with a wave vector in the $x$ direction. The propagation direction clearly has a $y$-component, which demonstrates that the propagation direction of spin-wave packets in chiral magnets can differ considerably from the direction of the wave packet's $k$-vector. The analytical calculation of the propagation, also shown in Fig.~\ref{fig:hall}, matches the simulated result perfectly.

\section{Spin-wave refraction in heterochiral magnetic films}

\subsection{A generalized Snell's law}

Spin waves reflect and/or refract at material boundaries. The momentum parallel to the interface should be conserved. Considering an interface along the $y$ direction, this translates to the constraint $k_{1,y}=k_{2,y}$, where the indices 1 and 2 denote the incident and refracted waves respectively. If the propagation direction is parallel to the $k$-vectors, the well known Snell's law applies: $k_1\sin\phi_1=k_2\sin\phi_2$ \footnote{All mentioned propagation angles are measured counterclockwise from the normal on the interface.}. If, however, the dispersion relation is asymmetric, then the propagation direction is not parallel to the wave vector and consequently Snell's law no longer describes the refraction of spin wave packets correctly.

In what follows, we examine the refraction of spin waves at interfaces between regions with different DMI strengths [$D(x<0)=D^{(1)}$ and $D(x>0)=D^{(2)}$] in three different ways. First, we employ full micromagnetic simulations using MuMax$^3$ \cite{Vansteenkiste2014}. Next, we demonstrate how to compute the refraction angle using the exact dispersion relation [Eq.~\eqref{eq:dispersion}]. Finally, we use the approximated dispersion relation [Eq.~\eqref{eq:approx}] to construct a generalized Snell's law which allows for analytical calculations of refraction angles as well as critical incident angles. The methods presented here can be easily extended to include changes in other material parameters as well. However, in order to capture the chiral effects solely and for the sake of clarity, we leave the other material parameters ($A$,$K_{\text{eff}}$,$M_{\text{s}}$) unchanged in regions where DMI is varied.

\begin{figure}[t]
    \centering
    \includegraphics{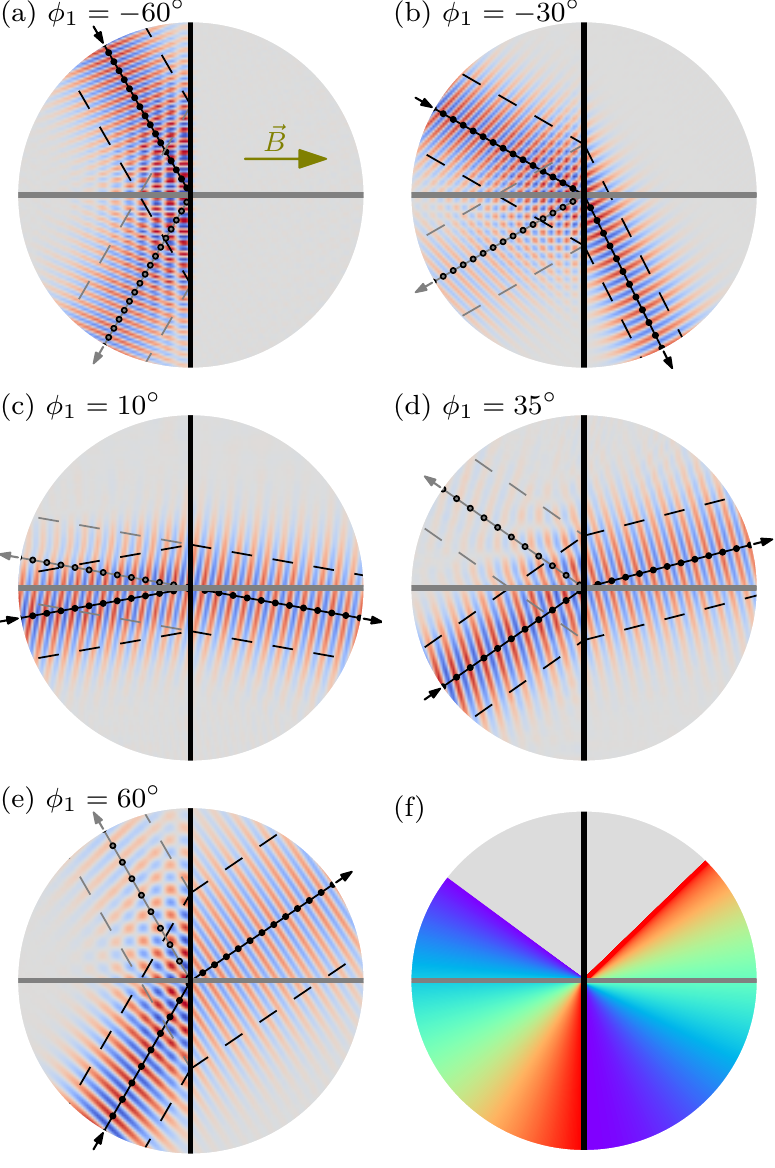} 
    \caption{(a-e) Refraction and reflection of wave packets under different incident angles $\phi_1$ at an interface with $D^{(1)}=0.9D_{\text{c}}$ on the left ($x<0$) and $D^{(2)}=0$ on the right ($x>0$), under an applied in-plane field $\vec{B}=1K/M_{\text{s}}\hat{e}_x$ perpendicular to the DMI interface. The contourplots show results of micromagnetic simulations with damping $\alpha=0.001$ and cell size $0.2\xi$. The wave packets are generated with a Gaussian antenna with frequency $\omega=1.5\omega_{\perp}$ and FWHM=$50\xi$ (dashed lines). The analytically predicted propagation direction and wavelength are depicted by solid lines and dots respectively. (f) Theoretical prediction of refraction for all possible incident angles. The direction of the incident and the corresponding refracted wave are plotted in same color. The gray region represents the range in which total reflection occurs.}
    \label{fig:refractionsim}
\end{figure}

The results of full micromagnetic simulations of wave packets incidental to a DMI interface are presented in Fig.~\ref{fig:refractionsim}(a-e). Qualitatively, they already show most of the interesting features of spin-wave refraction. The refraction is not symmetric for positive and negative incident angles. This is very clear when comparing the result for $\phi_1=-60^{\circ}$, for which there is total reflection, and $\phi_1=60^{\circ}$, for which there is noticeable transmission. Related to this, is the occurrence of negative refraction, visible in Fig.~\ref{fig:refractionsim}(c), where the incident and refracted waves are on the same side of the normal to the interface.

The interface makes the dispersion relation regionally dependent, for which term linear in $\vec{k}$ changes. To calculate the refraction angle for a given incident angle and frequency, we first compute $k_{1,y}$ by solving the dispersion relation [Eq.~\eqref{eq:dispersion}] for $\vec{k}_1$ in the left region, under the constraint that the direction of the group velocity $\vec{v}_{\text{g}}$ [Eq.~\eqref{eq:velocity}] corresponds to the given incident angle $\phi_1$. This can be done with a numerical self-consistent calculation. Imposing $k_{1,y}=k_{2,y}$, we solve the dispersion relation of the right region for $k_{2,x}$, by taking the real positive root of a fourth order equation. Once $\vec{k}_2$ is known, one can calculate the propagation direction with expression~\eqref{eq:velocity}. If the fourth order equation does not have positive roots, then there is total reflection at the interface. Note that we have neglected damping, higher-order deviations, and nonuniformities in the magnetization, such as the spin canting at the interface \cite{Mulkers2017}. However, Fig.~\ref{fig:refractionsim} shows that the calculated propagation directions, as well as the wavelengths, perfectly match the results of the simulations.

Using the circular isofrequency approximation of the dispersion relation, the condition $k_{1,y}=k_{2,y}$ can be rewritten in a generalized Snell's law:
\begin{equation}
    k^{(1)}_{\text{g}} \sin \phi_1 + k^{(1)}_{0,y}
    = k^{(2)}_{\text{g}} \sin \phi_2 + k^{(2)}_{0,y},
\end{equation}
enabling analytical calculation of the refraction angle~$\phi_2$ for a given incident angle~$\phi_1$. Yu {\it et al.}\cite{Yu2016} reported a similar generalized Snell's law for refraction of spin waves at a domain wall in a chiral magnet with an (atypical) in-plane easy anisotropy axis\footnote{Magnetization of chiral films is typically not in-plane due to an easy axis perpendicular to the film plane.}. At such (albeit uncharacteristic) domain walls, and in our case of chiral interfaces, the negative refraction and asymmetric Brewster angles occur due to the fact that the isofrequencies are shifted differently in $k$-space in the left and the right region, which is embodied in the generalized Snell's laws by the additional terms~$k^{(1)}_{0,y}$ and $k^{(2)}_{0,y}$. In contrast to the refraction at domain walls of Ref. \onlinecite{Yu2016}, there is no symmetry between the shifts in $k$-space in the left and right region in our case of a DMI interface. Furthermore, in our generalized Snell's law, $\vec{k}^{(1)}_0$ and $\vec{k}^{(2)}_0$ do not only depend on the DMI strengths, but can also be positioned in $k$ space at will by tuning the direction and magnitude of the in-plane bias field.

\begin{figure}[t]
    \centering
    \includegraphics{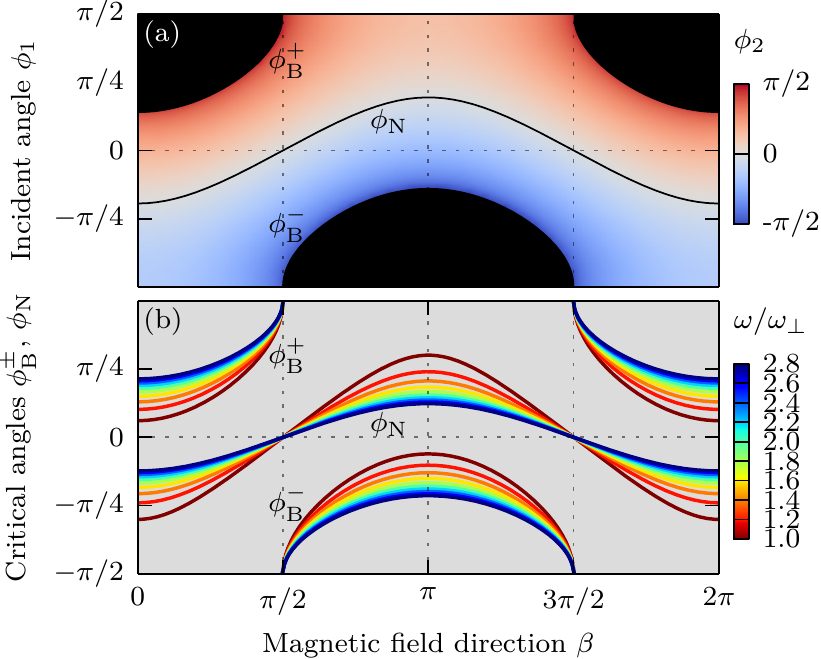} 
    \caption{(a) Refraction angle~$\phi_2$ of a wave packet with frequency~$\omega=1.5\omega_{\perp}$ at a DMI interface ($D^{(1)}=0$ and $D^{(2)}=0.9D_{\text{c}}$), as a function of the incident angle~$\phi_1$ and the direction of the applied field~$\vec{B} =1K/M_{\text{s}}(\cos\beta,\sin\beta,0)$. The black regions indicate total refraction, bounded by the Brewster angles~$\phi^{\pm}_{\text{B}}$, and the black line shows the critical negative refraction angle $\phi_{\text{N}}$. (b) The critical incident angles $\phi^{\pm}_{\text{B}}$ and $\phi_{\text{N}}$ as a function of the direction of the applied field (in-plane angle $\beta$) and frequency~$\omega$.}
    \label{fig:refraction}
\end{figure}

\subsection{Critical angles}

There are two different kinds of critical incident angles - the Brewster angles and the critical angle for negative refraction. The Brewster angles~$\phi_1=\phi^{\pm}_{\text{B}}$ can be calculated from the generalized Snell's law by imposing that the refracted wave is parallel to the interface ($\phi_2=\pm\pi/2$), as
\begin{equation}
    \phi^{\pm}_{\text{B}} = \arcsin \left( \frac{\pm k^{(2)}_{\text{g}} + k^{(2)}_{0,y} - k^{(1)}_{0,y}}{k^{(1)}_{\text{g}}} \right).
\end{equation}
The critical negative refraction angle is defined as the incident angle $\phi_1=\phi_{\text{N}}$ for which the refracted wave packet is orthogonal to the interface ($\phi_2=0$). Negative refraction occurs for incident angles between 0 and $\phi_{\text{N}}$. Using the generalized Snell's law, we obtain:
\begin{align}
    \phi_{\text{N}} &= \arcsin \left( \frac{k^{(2)}_{0,y} - k^{(1)}_{0,y}}{k^{(1)}_{\text{g}}} \right)  \nonumber \\
           &= \arcsin\left( \sqrt{\frac{\omega_{\perp}}{\omega-\omega^{1}_0}} \frac{2\sin\theta}{\pi}\frac{D^{(2)}-D^{(1)}}{D_{\text{c}}} \hat{e}_B\cdot\hat{e}_x \right).
\end{align}
Fig.~\ref{fig:refraction}(a) shows how the refraction angle and the critical angles depend on the incident angle~$\phi_1$ and the direction of the in-plane bias field (angle $\beta$). The asymmetry for positive and negative incident angles is clearly visible. Likewise, the two Brewster angles $\phi^{+}_{\text{B}}$ and $\phi^{-}_{\text{B}}$ are not equal. Fig.~\ref{fig:refraction}(b) shows the critical angles in function of the direction of the field and the frequency $\omega$. From this figure one can conclude that spin-wave packets with a low frequency refract more strongly than spin-wave packets with a high frequency.

\begin{figure}[t]
    \centering
    \includegraphics{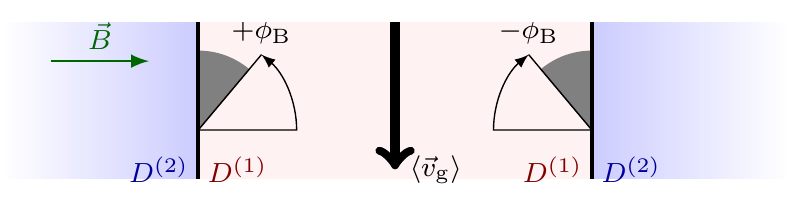} 
    \caption{A cartoon of a strip with DMI strength~$D^{(1)}$ within an extended ferromagnetic film with DMI strength $D^{(2)}>D^{(1)}$. Such a strip acts as a unidirectional spin wave guide when applying a magnetic field~$\vec{B}$ as shown. The Brewster angle $\phi_{\text{B}}$ delimits the angular section (grey) in which total reflection occurs. $\left<\vec{v}_{\text{g}}\right>$ is the average propagation direction of reflected waves.}
    \label{fig:waveguide}
\end{figure}

For given strengths of the applied field~$B$ and DMI, the range of total internal reflection is maximized when the in-plane field is perpendicular to the DMI interface. Note that in such a case there is no internal reflection for spin waves coming from the other side of the interface normal. Therefore, by adding a second parallel interface, one obtains an unidirectional spin-wave guide as illustrated schematically in Fig.~\ref{fig:waveguide}. Inverting the polarity of the applied magnetic field changes the direction of this spin-wave guide. This concept promotes heterochiral films as a broadly tunable platform for nanoengineered unidirectional spin-wave guides.

\section{Conclusions}

In summary, we have shown how the spin-wave dispersion relation in a chiral thin film with perpendicular anisotropy can be analyzed with a comprehensible geometrical representation, and derived a broadly-tunable Snell's law for a DMI interface, both checked against full-blown micromagnetic simulations. Bearing in mind the recent advances in direct imaging of incident, reflected and refracted spin waves in ferromagnetic films \cite{Stigloher2016}, and the emergent atomically-thin heterosystems where DMI can be spatially adjusted \cite{Woo2016,Moreau-Luchaire2016,Balk2017,Wells2017}, we expect our findings to inspire further theoretical and experimental work to explore full versatility of heterochiral ferromagnetic films for otherwise unattainable magnonic properties and devices.

\begin{acknowledgments}
This work was supported by the Research Foundation-Flanders (FWO-Vlaanderen) through Project No. G098917N.
\end{acknowledgments}

\appendix

\section{Derivation of the dispersion relation}

In this Appendix we provide the derivation of the dispersion relation for spin waves in a chiral ferromagnetic film with perpendicular easy anisotropy axis, subjected to an applied in-plane field~$B\hat{e}_y$ in the $y$ direction, within the micromagnetic framework. This dispersion relation can then easily be generalized for an arbitrary applied field direction in the $(x,y)$-plane.

The dynamics of magnetization $\vec{m}(x,y)$ is described by the Landau-Lifschitz-Gilbert equation. Here, we will only consider the precessional motion and neglect the damping term ($\alpha=0$). The equation of motion reads
\begin{equation}
    \dot{\vec{m}} = -\gamma {\vec{m}} \times {\vec{h}},
\end{equation}
with $\vec{h}$ the effective magnetic field which is related the functional derivative of the magnetic free energy~$E=\int\varepsilon \text{d} V$ with respect to the magnetization. For the energy density terms given in equations~\eqref{eq:exchange}-\eqref{eq:dmi}, the effective magnetic field reads
\begin{eqnarray}
    \vec{h} = -\frac{\delta E}{\delta\vec{M}} =&&  \frac{2A}{M_{\text{s}}} \Delta \vec{m} + \frac{2D}{M_{\text{s}}} \begin{pmatrix}  \partial_x m_z\\ \partial_y m_z \\ -\partial_x m_x - \partial_y m_y  \end{pmatrix}  \nonumber\\
    +&&  \frac{2K_{\text{eff}}}{M_{\text{s}}} m_z \hat{e}_z + B \hat{e}_y.
\end{eqnarray}

Consider the uniform equilibrium state $\vec{m}_0$. Due to the in plane field $B\hat{e}_y$ and the perpendicular anisotropy, the magnetization $\vec{m}_0$ will have a $z$ component as well as a $y$ component. The exact orientation of the magnetization $\vec{m}_0$ can be found easily by minimizing the free energy $E$ assuming a uniform magnetization:
\begin{equation}
    \vec{m}_0(x,y) = (0,\sin\theta,\cos\theta),
\end{equation}
with
\begin{equation}
    \theta =
    \begin{cases}
        \arcsin(M_{\text{s}} B/2K_{\text{eff}}), & \text{if } M_{\text{s}}B<2K_{\text{eff}},  \ \\
        \pi/2, & \text{if } M_{\text{s}} B \ge 2K_{\text{eff}}.
    \end{cases}
\end{equation}
Let us construct a new coordinate system for the magnetization $(\hat{e}_a,\hat{e}_b,\hat{e}_o)$ by rotating the coordinate system $(\hat{e}_x,\hat{e}_y,\hat{e}_z)$ around $\hat{e}_x$ over the angle $\theta$, making $\hat{e}_o$ and $\vec{m}_0$ parallel. The coordinate transformation is given by:
\begin{eqnarray}
    \hat{e}_a &=& \hat{e}_x, \nonumber\\
    \hat{e}_b &=& \cos\theta \hat{e}_y - \sin\theta\hat{e}_z,  \nonumber\\
    \hat{e}_o &=& \sin\theta \hat{e}_y + \cos\theta\hat{e}_z.
\end{eqnarray}

Expressing the effective field in the coordinate system $(\hat{e}_a,\hat{e}_b,\hat{e}_o)$ yields:
\begin{widetext}
\begin{align}
    \frac{M_{\text{s}} h_a}{2} &= A\Delta m_a -\sin\theta D \partial_x m_b + \cos\theta D \partial_xm_o,  \\
    \frac{M_{\text{s}} h_b}{2} &= A\Delta m_b +D\partial_y m_o + \sin\theta D \partial_x m_a + \cos\theta \frac{BM_{\text{s}}}{2} + \sin^2\theta K_{\text{eff}} m_b - \sin\theta\cos\theta K_{\text{eff}} m_o,  \\
    \frac{M_{\text{s}} h_o}{2} &= A\Delta m_o -D\partial_y m_b -\cos\theta D \partial_x m_a + \sin\theta \frac{BM_{\text{s}}}{2} + \cos^2\theta K_{\text{eff}} m_o - \sin\theta\cos\theta K_{\text{eff}} m_b.
\end{align}
\end{widetext}
Now we can study the time evolution of small deviations ($m_a\ll 1,m_b \ll 1, m_o\approx1$) from the equilibrium magnetization $\vec{m}_0$. For first-order deviations we obtain $\dot m_o = 0$ and
\begin{widetext}
\begin{align}
    \dot{m}_a &= -\gamma \left[m_bh_o-m_oh_b\right] \approx \frac{2\gamma}{M_{\text{s}}} (A\Delta-\cos^2\theta K_{\text{eff}} - \sin\theta \frac{BM_{\text{s}}}{2}+\sin^2\theta K_{\text{eff}} )m_b+\frac{2\gamma}{M_{\text{s}}} (\sin\theta D\partial_x)m_a, \\
    \dot{m}_b &= -\gamma \left[m_oh_a-m_ah_o\right] \approx-\frac{2\gamma}{M_{\text{s}}}( A\Delta-\sin\theta \frac{BM_{\text{s}}}{2}-\cos^2\theta K_{\text{eff}})m_a + \frac{2\gamma}{M_{\text{s}}} (\sin\theta D\partial_x)m_b.
\end{align}
\end{widetext}
The spin-wave dispersion relation can be calculated by filling in the plane waves $m_a \propto \exp i(\omega t - \vec{k}\cdot\vec{r})$ and $m_b \propto \exp i(\omega t - \vec{k}\cdot\vec{r})$, and solving the resulting system of equations, to obtain:
\begin{widetext}
\begin{equation}
    \frac{\omega}{\omega_{\perp}} =  \sqrt{\left(\xi^2k^2 + \cos^2\theta-\sin^2\theta +\sin\theta\frac{BM_{\text{s}}}{2K_{\text{eff}}} \right)\left(\xi^2k^2 + \cos^2\theta+\sin\theta\frac{BM_{\text{s}}}{2K_{\text{eff}}} \right)} - \sin\theta \frac{D}{\sqrt{AK_{\text{eff}}}} \xi k_x.
\end{equation}
\end{widetext}
Here we introduced the exchange length $\xi=\sqrt{A/K_{\text{eff}}}$, the critical DMI strength~$D_{\text{c}} = 4\sqrt{AK_{\text{eff}}}/\pi$, and the characteristic frequency $\omega_{\perp} = 2\gamma K_{\text{eff}}/M_{\text{s}}$.  The dispersion relation can be written in a simpler form if we define $\mathcal{B}=\max(1,M_{\text{s}}B/2K_{\text{eff}})$, as:
\begin{equation}
    \frac{\omega}{\omega_{\perp}}
    = \sqrt{\left(\xi^2k^2 + \mathcal{B} - \sin^2\theta\right)\left(\xi^2k^2+\mathcal{B}\right)} -\frac{4 \sin\theta}{\pi} \frac{D}{D_{\text{c}}} \xi k_x.
\end{equation}
Finally, we can generalize the dispersion relation for an arbitrary direction of applied field in the $(x,y)$-plane, as:
\begin{equation}
    \frac{\omega}{\omega_{\perp}}
    = \sqrt{\left(\xi^2k^2 + \mathcal{B} - \sin^2\theta\right)\left(\xi^2k^2+\mathcal{B}\right)} -2\xi^2\vec{k}\cdot \vec{k}_0,
\end{equation}
where
\begin{equation}
     \xi\vec{k}_0 = \frac{2 \sin\theta}{\pi}\frac{D}{D_{\text{c}}} (\hat{e}_B\times\hat{e}_z).
\end{equation}

\bibliography{main.auto}

\end{document}